\documentclass[final,letterpaper]{aipproc}
\layoutstyle{8x11double}
\usepackage{xspace}

\newcommand{\mathff}[1]{\mathrm{#1}}

\newcommand{\Msun}{{\ensuremath{\mathff{M}_{\odot}}}\xspace}

\def\gtaprx {\lower .1ex\hbox{\rlap{\raise .6ex\hbox{\hskip .3ex
        {\ifmmode{\scriptscriptstyle >}\else
                {$\scriptscriptstyle >$}\fi}}}
        \kern -.4ex{\ifmmode{\scriptscriptstyle \sim}\else
                {$\scriptscriptstyle\sim$}\fi}}}
\def\ltaprx {\lower .1ex\hbox{\rlap{\raise .6ex\hbox{\hskip .3ex
        {\ifmmode{\scriptscriptstyle <}\else
                {$\scriptscriptstyle <$}\fi}}}
        \kern -.4ex{\ifmmode{\scriptscriptstyle \sim}\else
                {$\scriptscriptstyle\sim$}\fi}}}

\begin{document}

\title[The Theory of Gamma-Ray Burst Central Engines]{The Central
Engines of Gamma-Ray Bursts}


\author{S. E. Woosley}{
  address={Department of Astronomy and Astrophysics, University of California, Santa Cruz, CA 95064, USA},
  email={woosley@ucolick.org},
  homepage={http://supersci.org}
}

\author{Weiqun Zhang}{
 address={Department of Astronomy and Astrophysics, University of California, Santa Cruz, CA 95064, USA},
  email={zhang@ucolick.org},
  homepage={http://supersci.org}
}

\author{A. Heger}{
  address={Enrico Fermi Institute,  University of Chicago, Chicago, IL 60637, USA},
  email={alex@ucolick.org},
  homepage={http://2sn.org},
}

% \copyrightholder{Acoustical Society of America}
\copyrightyear  {2001}

\begin{abstract}

Leading models for the ``central engine'' of long, soft gamma-ray
bursts (GRBs) are briefly reviewed with emphasis on the collapsar
model. Growing evidence supports the hypothesis that GRBs are a
supernova-like phenomenon occurring in star forming regions, differing
from ordinary supernovae in that a large fraction of their energy is
concentrated in highly relativistic jets. The possible progenitors and
physics of such explosions are discussed and the important role of the
interaction of the emerging relativistic jet with the collapsing star
is emphasized. This interaction may be responsible for most of the
time structure seen in long, soft GRBs. What we have called ``GRBs''
may actually be a diverse set of phenomena with a key parameter being
the angle at which the burst is observed. GRB 980425/SN 1988bw and the
recently discovered hard x-ray flashes may be examples of this
diversity.

\end{abstract}

\date{\today}

\maketitle

\section{Introduction}

Recent years have seen a welcome decline in estimates of the energy
that must be provided by the central engine of GRBs - from almost
10$^{54}$ erg to $\sim10^{52}$ erg (including the supernova that
accompanies the GRB). While retaining their title as the {\sl
brightest} explosions in the modern universe, current estimates of
kinetic and neutrino energies have demoted GRBs to merely comparable
to supernovae. This decrease has come about mostly because the beaming
long indicated by the theoretical models has been verified
experimentally. 

While the energy requirements on the model have become less
problematic, other demands have arisen. The successful model must not
only deliver a few times 10$^{51}$ erg in highly relativistic ejecta,
it must collimate those ejecta into a narrow outflow with typical
width 0.1 radian. Adherents of the internal shock model further need
the Lorentz factor to vary rapidly so that the efficiency for making
gamma-rays is not small. Diverse light curves should have a natural
explanation and the events must occur in star forming regions. At
least occasionally, GRBs should give be accompanied by Type I
supernova, by which we mean a supernova, without hydrogen, powered at
peak light by the decay of radioactive $^{56}$Ni and $^{56}$Co.

One model that shows promise in satisfying these constraints is the
{\sl collapsar} model. Collapsars occur naturally in star forming
regions, make bright Type I supernovae, offer a natural mechanism for
narrow jet collimation, and can explain the diverse light curves of
long GRBs.  They may also explain why GRBs have a nearly constant
total energy. The model also has its difficulties. The large requisite
angular momenta are difficult to achieve when current estimates of
magnetic torques and mass loss rates are included and the model also
offers no clear route to making short hard GRBs. However, unlike other
models, the collapsar model makes many testable predictions.

These predictions (and postdictions) include: a) association of GRBs
with massive stars and star formation; b) an increasing fraction of
GRBs at low metallicity (high redshift); c) collimation of the outflow
into a narrow jet with typical opening
angle 0.1 radian; d) an asymmetric Type I supernova accompanying every
GRB; e) an event rate of about 1\% that of ordinary supernovae (we
only see a few tenths of a percent of these as GRBs); f) a distribution
of circum-source matter consistent with the wind of a Wolf-Rayet star
at the end of its life; g) possible large production of $^{56}$Ni by
the disk of the accreting black hole; h) continuing energetic outflows
for a long time after the GRB; i) diverse phenomena seen at different
polar angles; and j) a jet whose temporal structure (and hence a GRB
whose light curve) is relatively insensitive to the central engine, but
set by the interaction of the jet with the overlying star.

Not all of these predictions are unique to the collapsar model. Any
model that produces the same relativistic outflow in the middle of a
massive helium star will give similar results.  In the end, details
like the duration of the event, the $^{56}$Ni production, and other
specific properties of the supernova will help to distinguish the
operation of an engine that, except for faint signals in neutrinos and
gravity waves, is not directly observable.

\section{Basic Collapsars}

Generically a collapsar is a rotating massive star, devoid of hydrogen
envelope, whose central regions collapse to a black hole surrounded by
an accretion disk \citep{Woo93, Mac99}. Accretion of at least a solar
mass through this disk produces outflows that are further collimated
by passage through the stellar mantle. These flows attain high Lorentz
factor as they emerge from the stellar surface and, after traversing
many stellar radii, produce a GRB and its afterglows by internal and
external shocks respectively.

There are actually several ways to make a collapsar and each is likely
to have different observational characteristics.

\begin{itemize}

\item 
{A standard (Type I) collapsar is one where the black hole forms
promptly in a helium core of approximately 15 to 40 \Msun. There never
is a successful outgoing shock after the iron core first collapses. A
massive, hot proto-neutron star briefly forms and radiates neutrinos,
but the neutrino flux is inadequate to halt the accretion. For iron
cores near 1.9 solar masses, as can occur in massive metal-deficient
stars \citep{Woo02}, and soft equations of state, eventual collapse to
a black hole is assured. For most other equations of state, collapse
to a black hole is also certain if the iron core accretes an
additional half-solar mass or so without launching an outgoing
shock. Such an occurrence seems very likely in helium cores of high
mass because of the rapid accretion that characterizes the first
second after core collapse \citep{Fry99a}. Such high mass helium cores
are more frequently realized in systems with low metallicity because
low metallicity inhibits mass loss. The mass threshold for forming a
black hole promptly is not known with certainty, but based upon two
dimensional models with crude neutrino physics is thought to be around
10 solar masses of helium. A lower limit is the 6 solar mass helium
core of SN 1987A (a main sequence star of $\sim$ 20 \Msun) which
certainly did not form a black hole promptly \citep{Bur88} because we
saw the neutrino signal continue for $\sim 10$ s.}

\item
{A variation on this theme is the ``Type II collapsar'' wherein the
black hole forms after some delay - typically a minute to an hour,
owing to the fallback of material that initially moves outwards, but
fails to achieve escape velocity \citep{Mac01}. The time scale for
such an event is set by the interval between the first outgoing shock
and the GRB event. Such an occurrence is again favored by massive
helium cores and might have occurred in SN 1987A \citep{Bro94}. The
binding energy rises rapidly above main sequence masses of 25 \Msun in
metal-poor stars \citep{Woo02}. Delayed black hole formation of this
sort seems unavoidable above some threshold mass that is probably
smaller than that required for Type I collapsars. There is
nucleosynthetic evidence in low metallicity stars in our own Galaxy
for a progenitor population where most of the iron failed to escape
the star \citep{Dep02}.

Type II collapsars should be common events, probably more frequent than
Type I. They are also capable of producing powerful jets that might
make gamma-ray bursts. Unfortunately their time scale may be, on the
average, too long for the typical long, soft bursts. If the
GRB-producing jet is launched within the first 100 s or so of the
initial supernova shock, it still emerges from the star before the
supernova shock has gotten to the surface, i.e., when the star is
still dense enough to provide collimation. Their accretion disks are
also not hot enough to be neutrino dominated and this may affect the
accretion efficiency \citep{Nar01}.}

\item
{A third variety of collapsar occurs for extremely massive
metal-deficient stars (above $\sim$300 \Msun) that may have existed in
the early universe \citep{Abe02,Fry01}. For non-rotating stars with
helium core masses above 137 \Msun (main sequence mass 280 \Msun), it
is known that a black hole forms after the pair instability is
encountered \citep{Heg02}. It is widely suspected that such massive
stars existed in abundance in the first generation after the Big Bang
at red shifts $\sim$5 - 20. For {\sl rotating} stars the mass limit
for black hole formation will be raised. The black hole that forms
here, about 100 \Msun, is more massive, than the several \Msun
characteristic of Type I and II collapsars, but the accretion rate is
also much higher, $\sim$10 \Msun s$^{-1}$, and the energy released may
also be much greater. The time scale in the lab frame for this
accretion is of order 20 s or so, not so different from GRBs. However,
one must dilate both the spectrum and the time scale according to the
red shift. The spectrum even in the lab frame is unknown, but if it
were similar to nearby GRBs, one might expect long, hard x-ray flashes
rather than classical GRBs.}

\end{itemize}

For Type I and II collapsars it is also essential that the star loses
its hydrogen envelope before death. No jet can penetrate the envelope
in less than the light crossing time which is typically 100 s for a
blue supergiant and 1000 s for a red one. After running into
1/$\Gamma$ of its rest mass, a ballistic jet loses its energy.

\section{Progenitors}

As Heger \& Woosley show elsewhere in these proceedings, a bare helium
star born (e.g., from a merger) with equatorial rotation 10\% of
Keplerian and low metallicity can retain enough angular momentum to
form a centrifugally supported disk around a central black hole of
$\sim3\,\Msun$.  Without magnetic fields, the angular momentum is
sufficient to form a Kerr black hole and support most of the star in
an accretion disk.  However, when an approximate treatment of angular
momentum transport by magnetic fields is included \citep{Spr01} along
with mass loss, the resulting rotation become too low to form
centrifugally supported disks in the inner part of the core. Even
though our knowledge of magnetic torques inside evolved massive stars
is still quite uncertain, this is a concern for the collapsar model.

The mass loss rate of Wolf-Rayet stars (WR-stars) during helium
burning (i.e., most of their lifetime) is observed to be large, but
has an uncertain dependence on the metallicity. A value of
$\sim$10$^{-5}$ \Msun y$^{-1}$ should be typical for collapsar
progenitors (Heger \& Woosley, this volume) implying a number density
(of {\sl helium} nuclei) outside the star $\sim 10^3 r^{-2}_{16}$
cm$^{-3}$, large compared with what is inferred from afterglows
\citep{Pan01}. This wind is clumpy and has an uncertain angular
distribution. Perhaps the polar region has a lower mass loss.

Because the mass loss carries away both mass and angular momentum, it
is detrimental to collapsar production.  To the extent that mass loss
is suppressed in such stars by low metallicity, one may expect the
fraction of stars that become GRBs to increase with redshift. Lower
mass loss rates for red supergiant stars with low metallicity also
raises the maximum mass of helium core that can result from the
evolution of single stars. For solar metallicity this number is about
12 \Msun \citep{Woo02}. For 1/4 solar metallicity, the number is
already considerably higher.

No observations constrain the mass loss rate of WR-stars during the
post-helium burning phases (100 - 1000 years), nor have the necessary
stability analyses been carried out to see if such stars are
stable. The loss rate could be quite high (or conceivably, even lower
if these stages {\sl are} stable). Unlike supergiants, the surface of
a WR-star remains in sonic communication with the central core up to
the last few minutes of core collapse. Strong acoustic waves from
pulsationally unstable oxygen and silicon burning stages could, in
principle, expel large quantities of material out to $\sim$10$^{15}$
cm (the escape speed times the duration of oxygen
burning). Alternatively, it is known that helium cores between about
40 and 65 \Msun encounter the pulsational pair instability 
\citep{Heg02}. Up to several solar masses can be ejected with
supernova-like energies from days to years prior to the collapse of
the core to a black hole. These are probably too massive to be common
GRBs, but could occur occasionally, especially at low metallicity.

\section{Collapsars - A Standard Bomb?}

Frail et al. \citep{Fra01} have suggested that GRBs are a standard
bomb in the sense that, within an order of magnitude, they all have
the same total kinetic energy, $\sim 3 \times 10^{51}$ erg.  This can
be understood in the collapsar model as an approximate constancy of
the total accreted mass and black hole mass. Collapsars of Type I have
an accretion rate of $\sim$0.1 \Msun s$^{-1}$ for about 20 s into a
black hole initially of about 3 \Msun. Slower accretion rates can
continue for a longer time, but the total energy in all cases is
limited by the fact that the jet explodes the star shutting off its
own accretion.  The gravitational binding energy outside the iron core
of a 15 \Msun helium star is $\sim 2 \times 10^{51}$ erg (contributing
in part to the difficulty of exploding these stars by ordinary
means). Some of this falls into the black hole, but the jet needs to
deposit at least 10$^{51}$ erg in the star simply to unbind it and
shut off the accretion.  Much more energetic jets will not be produced
because the mass in the disk is limited and the explosion shuts off
the flow.

\section{The Jet-Star Interaction}

The propagation of relativistic jets through the massive, partially
collapsed progenitors of collapsars has been considered in
refs. \citep{Alo00, Zha02} and the reader is referred to those papers
for extensive discussions (see also Zhang, Woosley, \& MacFadyen, this
volume). Regardless of the initial Lorentz factor and opening angle,
after a few seconds the jet inside the star is characterized by two
shocks, one at the leading ``head'' moving subrelativistically, and
another deeper in where the initial outflow runs into material piled
up behind the leading shock. Some of the material it runs into has
also slowed due to interaction with the jet walls. Only the initial
outflow deep within the star remembers the properties of the central
engine. In calculations this outflow is taken either to be born highly
relativistic ($\Gamma \sim 100$) or to have such large internal energy
that it becomes highly relativistic before going very far.

Between the two shocks is material with moderate Lorentz factor
($\Gamma \sim 10$) and large internal energy per baryon, roughly 10
$m_o c^2$. It is this material that, after exiting the star, makes
most of the prompt burst. The large Lorentz factors that characterize
GRBs ($\gtaprx 100$) are developed outside the star as expansion
converts internal energy back into highly relativistic motion. That is
conversion of internal energy in the moving frame give $\Gamma \sim
10$ which translates into $\Gamma \sim 200$ in the laboratory frame.
Considerable lateral expansion of the jet also occurs after exiting
the star. This is also true of the mildly relativistic cocoon of
matter surrounding the exiting jet (see also \citep{Mes01}).

\subsection{GRB Light Curves}

Within this context, all short time scale variability of the central
engine itself is washed out by the first shock. Variations of energy
input where the jet is born, do not manifest themselves in the GRB
light curve.

Still, GRB light curves are known to be diverse and complex. Where
does the time structure originate? We believe that it comes from the
jet-star interaction. Mixture of nearly stationary matter into the jet
by the (relativistic) Kelvin-Helmholtz instability can load the jet
with baryons and slow it down. A particularly large wavelength
instability can even temporarily block the flow giving bursts that
seem to turn on and off multiple times.

Modulating the jet in this fashion has two effects. First, it can give
rise to complex light curves even for a constant power input at the
base. Second, it can provide the variable Lorentz factor necessary for
the internal shock model.

\subsection{Luminosity-Variability-Angle Correlations}

The angle with which the jet emerges increases with time as the star
is blown aside. An observer situated at a relatively large angle may
not even see the GRB until it has been operating at smaller polar
angles for some time. If the jet power or its efficiency for
conversion to gamma-rays declines with time, it will appear less
luminous.  This effect will lead to an ``arrow-head'' structure for
the distribution of relativistic material in the jet. Salmonson \&
Galama \citep{Sal02} have discussed how such a structure leads to increased
break times (and thus larger inferred opening angles) for observers
off axis.

Further, if the central engine provides constant power, the energy per
unit area in the jet that emerges will be larger for more focused
jets. The focusing is assumed to be dependent on the structure of the
progenitor star.  Narrower cylindrical jets will have a larger ratio
of surface to volume and will experience more Kelvin-Helmholtz
instability, thus making them more variable.

Putting these effects together, one expects that jets with larger
opening angles will make less luminous, less variable GRBs. This is
apparently the case \citep{Rei01, Fra01}. 

\section{Supernovae}

One of the earliest predictions of the collapsar model was that each
GRB should be accompanied by a supernova-like display. The idea that a
black hole formation in a rotating helium star would make a supernova
of some sort was discussed by Bodenheimer \& Woosley
\citep{Bod83}. Woosley \citep{Woo93} extended the idea to GRB
production.  In what many regard as a gross understatement, these
early models were characterized as ``failed supernovae'' because the
usual mechanism for producing an outgoing shock and a successful
explosion (neutron star formation and neutrino emission) was assumed
to be inoperable. Calculations showing that the passage of a
relativistic jet through the star not only leads to collimation of the
jet, but explosion of the star were carried out by MacFadyen, Woosley,
\& Heger \citep{Mac99,Mac01} and Khokhlov et al. \citep{Kho99}. The
kinetic energy energy available to the explosion is approximately the
work the jet does prior to breaking out of the star \citep{Zha02}. For
a jet power of $\sim 5 \times 10^{50}$ erg s$^{-1}$ (both jets) and a
traversal time 5 - 10 s, this gives $\sim3 \times 10^{51}$ erg,
comparable to but somewhat greater than the kinetic energy of an
ordinary supernova. For a typical 10 s GRB this gives jet energies -
after break out - also of about $3 \times 10^{51}$ erg per jet,
similar to what is inferred from afterglows and a kinetic energy
conversion of 20\%. Of course the supernova is initially grossly
asymmetric and one might infer a much more energetic explosion viewing
the supernova along the jet axis.

Lacking a hydrogen envelope, the supernova will be Type Ib or Ic with
an optical luminosity given entirely by the yield of $^{56}$Ni. It is
not generally appreciated how poorly determined this yield is in most
GRB models. In ordinary (spherically symmetric) supernovae the
iron-group yield (mostly $^{56}$Ni) is set by the amount of ejected
material that experiences explosion temperatures in excess of $5
\times 10^9$ K. This in turn is given by the strength of the explosion
and the density structure at the edge of the collapsing iron core.  In
Type I collapsars however, the material that would have become
$^{56}$Ni falls into the black hole. In Type II collapsars some
$^{56}$Ni is ejected, but not as much as in an explosion that left a
neutron star. Depending on the fallback mass, most of the $^{56}$Ni
reimplodes. The jet itself subtends a small solid angle and carries a
small, albeit very energetic mass. It cannot propagate outwards until
the mass flux inwards at the pole has declined, i.e., the density has
gone down. This makes it hard for the jet itself to synthesize much
$^{56}$Ni. How then is the supernova visible?

We think that the $^{56}$Ni may be made not by the jet, but by the
disk wind (\citep{Mac99,Nar01} and MacFadyen, this volume).  In the
parlance of Narayan et al., it could be that at late times (after
$\sim$10 s), a neutrino-dominated accretion disk (NDAF) switches to a
convection dominated accretion disk (CDAF) with a large fraction of
the mass flow being ejected. On the other hand, MacFadyen and Woosley
found considerable mass outflow even from NDAFs.  We postulate that a
certain fraction of the accreting matter - composed initially of
nucleons or iron group elements - is ejected at high velocity
($\sim$0.1 c) by the accretion disk. If the energy of the burst is
given by the amount of material accreted and the outflow fraction is
constant, the brightness of the supernova could correlate with the
energy of the GRB. This simple relation is could be complicated by
uncertain efficiency factors for converting disk energy into jet
energy and mass loss from the disk. If the accretion rate is very high
and the disk viscosity quite low, the density in the disk may be so
high that electron capture must be considered.

In any case, it is quite possible to get a highly variable amount of
$^{56}$Ni, and therefore a variable luminosity for the peak of the
supernova. SN 1998bw may not be a standard candle.  Observationally,
besides SN 1998bw, evidence for supernova-like light curves, color,
and time history has been found in at least three GRBs: GRB 011121
\citep{Blo02}; GRB 980326 \citep{Blo99}, and GRB 970228 \citep{Rei99,
Gal00}. What we would all like to see is the spectrum of a putative
supernova accompanying a cosmologically distant GRB.

\section{Alternate Models}

\subsection{Merging Neutron Stars}

The principal alternative model to the collapsar remains the merging
neutron star pair or neutron star - black hole pair discussed
elsewhere in this proceedings. These have the admirable properties of
being associated with events that are known to occur in nature and
have sufficient angular momentum to form an accretion disk around the
black hole after the merger. An energy of $3 \times 10^{51}$ erg in
relativistic ejecta is more challenging for these models than some
others, but easily within reach of those employing
magnetohydrodynamics (MHD) to extract black hole rotational energy or
disk binding energy. However, even though a few of these might happen
in star-forming regions, the vast majority are expected to occur
outside \citep{Fry99b}. It may also be difficult for merging neutron
stars to collimate their outflows within 0.1 radians, at least in
those versions where neutrino transport produces the jet. Given the
difficulty the collapsar model has in making short hard bursts, we
continue to associate compact mergers with this subclass
\citep{Fry99b}.

Other popular models for the long, soft GRBs associated with massive star
death involve either the delayed production of a black hole or the 
prompt production of a very magnetic, rapidly rotating neutron star. 

\subsection{``Supranovae''}

It has been suggested by Vietri\& Stella \citep{Vie98,Vie99} and
others that GRBs may result from the delayed implosion of rapidly
rotating neutron stars to black holes. The neutron star is
``supramassive'' in the sense that without rotation, it would
collapse, but with rotation, collapse is delayed until angular
momentum is lost.  The momentum can be lost by gravitational radiation
and by magnetic field torques.  Vietri and Stella assume that the
usual pulsar formula holds and, for a field of 10$^{12}$ gauss, a
delay of order years (depending on the field radius and mass) is
expected. When the centrifugal support becomes sufficiently weak, the
star experiences a period of runaway deformation and gravitational
radiation before collapsing into a black hole. It is assumed that
$\sim$0.1 \Msun is left behind in a disk which accretes and powers the
burst in a manner analogous to the merging neutron star model.

The model has several advantages. It, as well as the collapsar model
that it in some ways resembles, predicts an association of GRBs with
massive stars and supernovae. Moreover it produces a large amount of
material enriched in heavy elements located sufficiently far from the
GRB as not to obscure it. The irradiation of this material by the
burst or afterglow can produce x-ray emission lines as have been
reported in several bursts \citep{Pir99,Pir00,Ree02}.

However, the supranova model also has some difficulties (see also
\citep{Mcl02}). First, it may take fine tuning to produce a GRB days
to years after the neutron star is born. Shapiro \citep{Sha00} has
shown that neutron stars requiring differential rotation for their
support will collapse in only a few minutes.  The requirement of rigid
rotation reduces the range of masses that can be supported by rotation
to, at most, $\sim$20\% above the non-rotating limit
\citep{Sha00,Sal94}.  Small changes in the angular momentum and mass
cause large variations in the delay between the supernova and
GRB. This is because the pulsar radiation formula employed depends on
$\omega^{-4}$ and the critical angular momentum where collapse ensues
depends on the excess mass above the non-rotating limit. That the
combination would have been just right in a (randomly selected) event
like GRB 011211 \citep{Ree02} to give a delay of a few days is highly
constraining.

Second, the supernova had best happen years and not days
before the GRB (in conflict with Reeves et al). Supernovae are
optically thick to gamma-rays until well after their optical peak,
that is, at least a month even for Type I. Nor can any relativistic
jet penetrate an object whose light crossing time is well in excess of
the duration of the central engine ($\sim$10$^{12}$ cm). Supranovae
that are younger might make lines, but they don't make GRBs. Third,
the very success of the collapsar model in producing a collimated jet
with opening angle near 5 degrees must be held in its favor.  This
collimation, as well as the time structure in the GRB light curve
require a high pressure stellar mantle to be present when the black
hole launches its jet \citep{Zha02}. Finally, the timing of supernovae
seen in conjunction with GRBs demands a simultaneous explosion. The
optical maximum of a Type I supernova of any subclass occurs a few
weeks after explosion.  This time modulated by the redshift is
consistent with SN 1998bw/GRB 980425 and with supernovae seen in the
tails of the optical afterglows of several other GRBs.

The collapsar gets around these restrictions by producing a jet that
exits the star while the central engine is still on and making a
supernova nearly simultaneously.  Lines might be energized by a
continuation of the same jet at late times (see ``Post-Burst
Phenomenology''). In the event that it proves necessary to eject
appreciable matter just prior to the GRB, one may want to consider
pulsationally driven mass loss (see ``Progenitors'').

\subsection{``Magnetar Model''}

Another model, championed most recently by Wheeler et al
\citep{Whe00}, is the ``super-magnetar'' model (see also
\citep{Uso92}). As usual, the iron core of a massive star collapses to
a neutron star.  For whatever reasons, unusually high angular momentum
perhaps, the neutron star acquires at birth an extremely powerful
magnetic field, $10^{15}$ - 10$^{17}$ gauss. If the neutron star
additionally rotates with a period of a ms or so, up to 10$^{52}$ erg
in rotational energy can be extracted on a GRB time scale by a
variation of the pulsar mechanism. This model has the attractive
features of being associated with massive stars, making a supernova as
well as a GRB, and utilizing an object, the magnetar, that is
implicated in other phenomena - soft gamma-ray repeaters and anomalous
x-ray pulsars. It has the unattractive feature of invoking the
magnetar fully formed in the middle of a star in the process of
collapsing without consideration of the effects of neutrinos or rapid
accretion. The star does not have time to develop a deformed geometry
or disk that might help to collimate jets. To break the symmetry,
Wheeler et al invoke the operation of a prior LeBlanc-Wilson
\citep{Leb70} jet to ``weaken'' the confinement of the radiation
bubble along the rotational axis. Numerical models to give substance
to this scenario are needed (though see \citep{Whe02}).

\section{GRBs - A Unified Model}

According to the ``Unified Model'' for active galactic nuclei (e.g.,
\citep{Ant93}, one sees a variety of phenomena depending upon the
angle at which the source is viewed. These range from tremendously
luminous blazars, thought to be jets seen on axis, to narrow line
radio galaxies and Type 2 Seyferts thought to be similar sources seen
edge on. Given that an accreting black hole and relativistic jet may
be involved in both, it is natural to seek analogies with GRBs.

In the equatorial plane of a collapsar, probably little more is seen
than an extraordinary supernova that, were it close enough to observe,
would be an exceptionally bright radio source. Roughly 1\% of all
supernovae might be of this variety - perhaps a larger fraction at
high redshift. Given that we have seen $\sim$1000 relatively nearby
supernovae, it would not be surprising to find a few in the cataloged
sample. They would be of Type Ib/c and perhaps extraordinarily
energetic.  SN 1998bw could be a prototype, but without the high
velocities that come from observing the event at high latitude.  There
are indications that a few of these may have been seen. Besides SN
1998bw there are SN 1997ef and 1997ey \citep{Nak01}, and perhaps SN
2002ap. These supernovae, all of Type I, are characterized by a large
inferred kinetic energy (at least for the equivalent isotropic
explosion) and a variable, but occasionally large mass of
$^{56}$Ni. Very high velocity intermediate mass elements were also
seen in SN 1998bw \citep{Pat01}.  Completing the connection from the
other end, there are also an increasing number of GRBs which show
evidence for supernova-like activity in the tail of their optical
afterglow, most recently in GRB 011211 \citep{Blo02}.

Perhaps the most interesting phenomena are those at intermediate
angles. The models clearly show, and nature generally demands that the
edges of jets are not discontinuous surfaces. Moving off axis, one
expects and calculates a smooth decline in the Lorentz factor and
energy of relativistic ejecta. These low energy wings with moderate
Lorentz factor come about in three ways \citep{Zha02}. First, the jet
that breaks out still has a lot of internal energy. Expansion of this
material in the comoving frame leads to a broadening of the jet. Some
of the material is even {\sl decelerated} by expansion pushing back
towards the origin. As a result a small amount of material with low
energy ends up moving with intermediate Lorentz factors - say 10 - 30
and at angles up to several times that of the main GRB-producing
jet. Second, as the star explodes from around the jet, the emerging
beam opens up. At late times the outflow continues (see ``Post-burst
Phenomenology''), but with decreased power. Third, the jet is
surrounded by a hot mildly relativistic cocoon. This material has low
energy, but can expand to large angles. As a consequence of the
spreading of the jet, a large region of the sky, much larger than that
which sees the main GRB, will see a hard transient with less power,
lower Lorentz factor, and perhaps coming from an external shock
instead of internal ones.

We have speculated for some time now \citep{Woo98,Woo99,Woo00,Woo01}
that these ordinary bursts seen off axis might appear as hard x-ray
transients of one sort or another. We have identified them with GRB
980425 and with the class of hard x-ray flashes reported by Heise et
al. \citep{Hei01}. We do not say that these are ordinary high $\Gamma$
jets seen just beyond a sharp edge \citep{Iok01}. The events are made
by matter moving towards us.

In the particular case of GRB 980425, it is important to know if the
optical afterglow of a {\sl normal} GRB that was {\sl not} directed at
us would have had an observable effect on the supernova light curve.
After all, the optical afterglow is not nearly so beamed as the GRB
and might be visible at lower latitude.  However, Granot et
al. \citep{Gra02} show that the afterglow would be invisible at the
time the supernova was studied provided that the polar angle to our
line of sight is greater than about 3 or 4 times that of the main GRB.
This does raise the interesting possibility though that some future
event might show the supernova and afterglow more nearly balanced in a
``soft'' relatively faint GRB.

\section{Post-burst Phenomenology}

After the main burst is over, accretion continues at a decaying
rate. The lateral shock launched by the jet starts at the pole and
wraps around the star, but does not reach into the origin at the
equator (one may envision an angle-dependent ``mass
cut''). Consequently, some reservoir remains to be accreted at late
time. This accretion occurs at a rate given by the viscosity of the
residual disk and the free fall time of material farther out not
ejected in the supernova. MacFadyen, Woosley, \& Heger \citep{Mac01}
and Chevalier \citep{Che89} estimate the accretion rate from fall back
to be $\sim3 \times 10^{-6} t_4^{5/3}$ \Msun s$^{-1}$. Here $t_4$ is
the elapsed time since core collapse in units of 10$^4$ s. Given the
slow rate, the disk that forms is not neutrino dominated and there may
be considerable high velocity flow from its surface (MacFadyen, this
volume; \citep{Mac99,Nar01}).  The outflow will still be jet-like in
nature since the equatorial plane is blocked by the disk and its
energy will be $\sim 5 \times 10^{46} t_4^{-5/3} \epsilon_{01}$ erg \
s$^{-1}$ where $\epsilon_{01}$ is the efficiency for converting rest
mass into measured in percent. This is comparable to the energy in
x-ray afterglows and might be important for producing the emission
lines reported in some bursts \citep{Ree00,Mcl02} and for providing an
extended tail of hard emission in the GRB itself.

As a consequence of this continuing outflow, the polar regions of the
supernova made by the GRB remain evacuated and the photosphere of the
object resembles an ellipse seen along its major axis but with conical
sections removed along the axis. An observer can see deeper into the
explosion than they could have without the operation of the jet's
``afterburner''.

\section{Conclusions}

The collapsar model is able to explain many of the observed
characteristics of GRBs. Here we have explored some of its predictions
(enumerated in the ``Introduction''). Probably the greatest challenges
facing the model today are not the large energy associated with GRBs,
or even the relativistic collimated flow. They are an understanding of
how the necessary angular momentum comes about in the precollapse star
- presumably by the special circumstances that make GRBs rare compared
with supernovae - and of how accretion energy in the disk is
transformed into jets. The former is a problem we share with competing
models for GRBs like the supranova and millisecond magnetar models;
the latter is also a long standing obstacle in understanding
AGNs. There is hope that numerical simulation might address both in a
few years.

We have described a ``unified theory of GRBs'' in which diverse
phenomena are expected depending upon the angle at which a standard
model for the explosion is viewed. In this theory the Lorentz factor
and the energy of relativistic ejecta vary both with polar angle and
with time. This paradigm is similar to the unified theory of
AGNs. Some of the phenomena it predicts, like hard x-ray flashes are
just now being discovered. Others like the long duration gamma-ray
transients expected from Type II and III collapsars may await
discovery.

In the collapsar model, the light curves of GRBs reflect more the
interaction of the jet with the star as it emerges than time
variability of the central engine itself. We have mentioned how
correlations in break times, luminosity, and opening angle might come
about and discussed some of the special properties of the accompanying
supernova. 

In the near future we hope to carry out the next steps in realistic
collapsar simulation - special relativistic studies (in three
dimensions) of jet propagation inside the star and longer time scale
calculations of the supernova it produces.

\begin{theacknowledgments}

This work has been supported by the HETE-2 grant (MIT-SC-292701), the
NASA Theory Program (NAG5-8128), and by the Scientific Discovery
Through Advanced Computing (SciDAC) program of the DOE
(DE-FC02-01ER41176). Alex Heger was partly supported by the Alexander
von Humboldt Society (FLF-1065004). We are grateful for helpful
conversations with and calculations by Andrew MacFadyen regarding the
nature of collapsars.

\end{theacknowledgments}

\end{document}